# JU_KS_Group@FIRE 2016: Consumer Health Information Search

Kamal Sarkar
Dept. of Computer Sc. & Engg.
Jadavpur University
Kolkata, WB 700032
jukamal2001@yahoo.com

Debanjan Das
Dept. of Computer Sc. & Engg.
Jadavpur University
Kolkata, WB 700032
dasdebanjan624@gmail.com

Indra Banerjee
Dept. of Computer Sc. & Engg.
Jadavpur University
Kolkata, WB 700032
ardnibanerjee@gmail.com

Mamta Kumari
Dept. of Computer Sc. & Engg.
Jadavpur University
Kolkata, WB 700032
mamta.mk222@gmail.com

Prasenjit Biswas
Dept. of Computer Sc. & Engg.
Jadavpur University
Kolkata, WB 700032
p.biswas.ju94@gmail.com

## ABSTRACT
In this paper, we describe the methodology used and the results obtained by us for completing the tasks given under the shared task on *Consumer Health Information Search* (CHIS) collocated with the Forum for Information Retrieval Evaluation (FIRE) 2016, ISI Kolkata. The shared task consists of two sub-tasks – (1) task1: given a query and a document/set of documents associated with that query, the task is to classify the sentences in the document as relevant to the query or not and (2) task 2: the relevant sentences need to be further classified as supporting the claim made in the query, or opposing the claim made in the query. We have participated in both the sub-tasks. The percentage accuracy obtained by our developed system for task1 was 73.39 which is third highest among the 9 teams participated in the shared task.

## Categories and Subject Descriptors
H.1.2 [Information Systems]: User/Machine Systems – human factors, human information processing.

## 1. INTRODUCTION
### 1.1 Our Motivation
A large number of websites provide health related information [1][2]. Consumer use of the Internet for seeking health information is rapidly growing [3]. By 1997, nearly half of Internet users in the US had sought health information [4]. Expressed in raw numbers, an estimated 18 million adults in the US sought health information online in 1998. The majority of consumers seek for themselves health information related to diseases for consultation with their physicians [5] [6]. Information found trough search on the web may influence medical decision making and help consumers to manage their own care [7]. The most common topics which are searched on the web are the leading causes of death (heart disease and cancer) and Children health.

Information access mechanisms for factual health information retrieval have matured considerably, with search engines providing Fact Checked Health Knowledge Graph search results to factual health queries. It is pretty straightforward to get an answer to the query "what are the symptoms of Diabetes" from these search engines [8][9][10]. But the most general purpose search engines can hardly find the answers of the complex health search queries which do not have a single definitive answer and whose answers have multiple perspectives. There may have a search queries for which there are a large number of search results reflecting the different perspectives and view-points in favor or against the query.

The term "Consumer Health Information Search" (CHIS) has been used by the organizers of the shared task on Consumer Health Information Search @FIRE 2016 to denote such information retrieval search tasks for which there are no "Single Correct Answer(s)" and instead, multiple and diverse perspectives/points of view, which very often are contradictory in nature, are available on the web regarding the queried information[1].

### 1.2 Problem Statement
The shared task on Consumer Health Information Search @FIRE 2016 has the following two sub-tasks:

A) Task 1- Given a CHIS query and a document/set of documents associated with that query, the task given was to classify the sentences in the document as relevant to the query or not. Relevant sentences in the document being those which are useful in providing the answer to the query.

B) Task 2- These relevant sentences had to be further classified as supporting the claim made in the query or opposing it.

*1.2.1 Examples*

E.g. Query - Are e-cigarettes safer than normal cigarettes?

S1: Because some research has suggested that the levels of most toxicants in vapor are lower than the levels in smoke, e-cigarettes have been deemed to be safer than regular cigarettes.

A) Relevant, B) Support

---

[1] https://sites.google.com/site/multiperspectivehealthqa/

S2: David Peyton, a chemistry professor at Portland State University who helped conduct the research, says that the type of formaldehyde generated by e-cigarettes could increase the likelihood it would get deposited in the lung, leading to lung cancer.

A)Relevant, B) oppose

S3: Harvey Simon, MD, Harvard Health Editor, expressed concern that the nicotine amounts in e-cigarettes can vary significantly.

A) Irrelevant, B) Neutral

## 2. METHODOLOGY
## 2.1 Description

For both the tasks-Task 1 and Task 2, we have used support vector machines (SVM) as the classifier, but the feature sets for the task1 and task2 were different. We discuss the feature sets used for task1 and task 2 in sub section 2.1.1 and sub-section 2.1.3 respectively.

### 2.1.1 Our Used Features for Task1

For the task 1, we were given by the organizers of the shared task a set of excel files where the heading of each excel file was a user query. Each excel file contained a set of sentences that were labeled as relevant or not relevant to the user query. The sentences in these given training excel files were already labeled as 'relevant' or 'irrelevant'. We took each and every sentence from each excel file and pair it with the corresponding query, examined them, and calculated a set of five features discussed in this sub section.

*2.1.1.1 Exact Matching:* We matched each sentences with the user given query, word by word, and calculated the similarity between the user query and the current sentence in the excel file; e.g.

Let the user query be "Ram is a good boy" and the current sentence be "Shyam is a bad boy". Between the user query and the current sentence there are three words which are exactly matching, i.e. "is", "a" & "boy". Now the similarity between these two strings is given as;

Similarity = {2 * (No. of Common Words)} / {(No. of words in user query) + (No. Of words in the current sentence)}     --- (i)

where, no. of Common Words = Number of words common to both the user query and the current sentence.

*2.1.1.2 Stemmed Word Matching:* We stemmed both the user query and the current sentence using a stemming tool available in Python programming language. Stemming normalizes a word by cutting out the excess part of a word due to pluralization, or if the word is an adverb; e.g. mangoes → mango, highly → high etc. After stemming we again calculated the similarity between both the strings using equation (i).

*2.1.1.3 Noun matching:* We found, on a perusal of initial sample data, that the nouns present in each sentence largely influenced whether a search result was relevant or irrelevant to the user query. So we isolated the nouns present in the user query, searched whether any of these nouns were matching with any word present in current sentence, and by this process we found out the number of nouns present in the current sentence that were exactly matching the nouns present in the user query. We calculated the noun matching similarity using the following formula;

Noun Similarity = (No. of nouns that are exactly matching with the nouns in the query) / (No. of nouns present in the user query)
     --- (ii)

*2.1.1.4 Neighborhood Matching:* There were some words present in the sentences which were not matching exactly with the words of user query, but they are semantically similar with the user query words; e.g.

Let 'skin cancer' be present in the user query and 'melanoma' be present in the current sentence. Both words are spelt differently but their meanings are similar, i.e. they are meaningfully similar.

To check whether the words were equivalent or not, we took each word from the current sentence, searched it in our self-made Wikipedia Dictionary [11], and extracted the first three sentences describing that word's meaning. We then matched the user query words with the words present in the extracted sentences, and if the word is present, we consider it as a match and, finally we calculate the similarity again between the user query and the current sentence using equation (i).

We create the Wikipedia Dictionary by saving words along with their meanings, which were extracted from Wikipedia. We use our developed python script for creating this dictionary.

*2.1.1.5 COSINE Similarity:*

We represent both the query and a sentence using bag-of-words model and each query as well as the sentence is represented as vector. The component of each vector is TFIDF weight of a word $t$ which is calculated as follows:

IDF (t) = log(N/DF)

Where N= Total number of sentences and DF= Number of sentences with word 't' in it

TF(t) = (Number of times word 't' appears in a sentence) / (Total number of words in the sentence)

After calculating the vectors for the query and the sentence, the cosine similarity between the query vector and the sentence vector is calculated. The cosine similarity value is used as one of feature values for relevance checking.

### 2.1.2 Search as Classification

For task 1, we represent each training sentence as vector of five feature values mentioned above and label each vector as "relevant" or "not relevant". With this labeled training data, we train the support vector machines (SVM). For SVM, we have used SVC tool available in Python scikit learn and a model is generated. Since no development set was available, for parameter tuning, we split the training data into two parts-(1) the first part contains 60% of the training data and second part contain 40% of the training data. We train SVM with the 60% of the training data and then we test the obtained model on the remaining part of the training data. Thus we tune the parameters to obtain the best parameter settings. Finally, we obtain the best results with the settings where the cost parameter *C* set to $10^7$, *gamma* set to 0.006 and *kernel* set to "poly".

Like training data, we represent the unlabeled test data released by the organizers of the shared task in the similar way using the five features mentioned in sub-section 2.1.1, and then submit it to the trained classifier. The classifier, using its knowledge from previous training data, predicts the labels for each of the sentences present in test data.

### 2.1.3 Our Used Features for Task2

After relevancy checking (Task 1), Task 2 is carried out. By task 1, all the sentences in the excel file are divided into two classes; (a) relevant and (b) irrelevant. Now the task is to determine whether a relevant sentence was supporting the user query, opposing the user query, or neutral with regard to the user query. For this task we again calculated a set of N+4 features, where N = number of distinct words present in the entire training files. Here the feature set includes N number of distinct unigrams present in the training data and four other features discussed in the following sub-sections.

*2.1.3.1 Number of Positive Words:* We calculated the number of positive words that were present in each sentence of the excel file. We recognized the positive words from a particular sentence by using a Python package called SentiWordNet[2].

*2.1.3.2 Number of Negative Words:* We calculated the number of negative words that were present in each sentence of the excel file. We recognized the negative words from a particular sentence by using a Python package called SentiWordNet.

*2.1.3.3 Number of Neutral Words:* We had already found out the positive and negative words for a particular sentence, so the words that were neither negative nor positive were classified as neutral words and their occurrence in the current sentence was counted.

*2.1.3.4 Relevant or Irrelevant:* In Task-1 we have already labeled each sentence to be either relevant or irrelevant. We took this label into consideration for this task. This was a binary feature as the current sentence could either be relevant or irrelevant.

*2.1.3.5 'N' Features:* we represent each sentence as a bag-of-words model. According to vector space model, a sentence is represented as N-dimensional vectors where N is the distinct number of unigrams present in the training data. Weight of a word used as the component of a vector is calculated using TFIDF formula.

### 2.1.4 Sentiment Classification

We represent each sentence in the excel file as a vector using the above mentioned N+4 features and label each vector with the label of the corresponding training sentence. The label can be one of three types- "Support", "Oppose" and "Neutral". Finally, we submit labeled vectors to the SVM classifier as specified in the Task-1 and trained it using them. The model is generated after training. Like the task 1, we also we split the training data into two parts-(1) the first part contains 60% of the training data, which is used to develop the initial model and (2) the remaining 40% of the training data is used to test the model while tuning the parameters. After tuning the parameters of SVC tool available in Python scikit learn, we obtain the best model with the cost parameter *C* set to $10^7$, *gamma* set to 0.005 and *kernel* set to "rbf".

We also represent unlabeled test data released by the organizers for the task 2 as the vectors using the same feature set consisting of N+4 features and submit them to the trained model which in turn predicts label 'supporting'/'opposing'/'neutral' for each sentence present in the test excel file.

## 2.2 Architecture

The architecture of our developed system used for task 1 and task 2 are shown in Figure 1 and Figure 2 respectively. For both the systems, the important modules are feature extraction and classifier. For the task 1, we have 5 features discussed in the earlier sections and for task 2, we have used N + 4 features which are also discussed in the earlier sections.

For task 1, after feature extraction from each query-sentence pairs, each sentence is represented as a vector which is labeled with the label of the corresponding training sentence. Then the labeled vectors are given to the classifier to produce a model. Finally the learned model is used to determine the relevancy of the test sentences given a query. For task 2, we extract features from the sentences and sentences are represented as the vectors labeled with one of the categories-"oppose", "support" and "neutral". The classifier is trained with the labeled training pattern vectors and the learned model is used to classify the test sentences into one of categories-"oppose", "support" and "neutral".

## 3. DATA SETS, RESULTS, EVALUATION

### 3.1 Data Sets

For the training data, we were given five user queries along with number of sentences per query [12].

- "does_sun_exposure_cause_skin_cancer"    -- 68 sentences
- "e – cigarettes"    -- 83 sentences
- "HRT_cause_cancer"    -- 61 sentences
- "MMR_vaccine_lead_to_autism"    -- 71 sentences
- "vitamin_C_common_cold"    -- 65 sentences

A total of 348 sentences were present in the training data set.

For the test data, the queries were the same as the training data and the number of unlabeled sentences per query given was as follows.

- "does_sun_exposure_cause_skin_cancer"    -- 342 sentences
- "e – cigarettes"    -- 414 sentences
- "HRT_cause_cancer"    -- 260 sentences
- "MMR_vaccine_lead_to_autism"    -- 279 sentences
- "vitamin_C_common_cold"    -- 247 sentences

A total 1542 sentences were present in the test data set

### 3.2 Results

We developed our systems for both task 1 and task 2 using the training data [12] supplied to us by the organizers of the contest.

---

[2] http://www.nltk.org/howto/sentiwordnet.html

http://www.nltk.org/_modules/nltk/corpus/reader/sentiwordnet.html

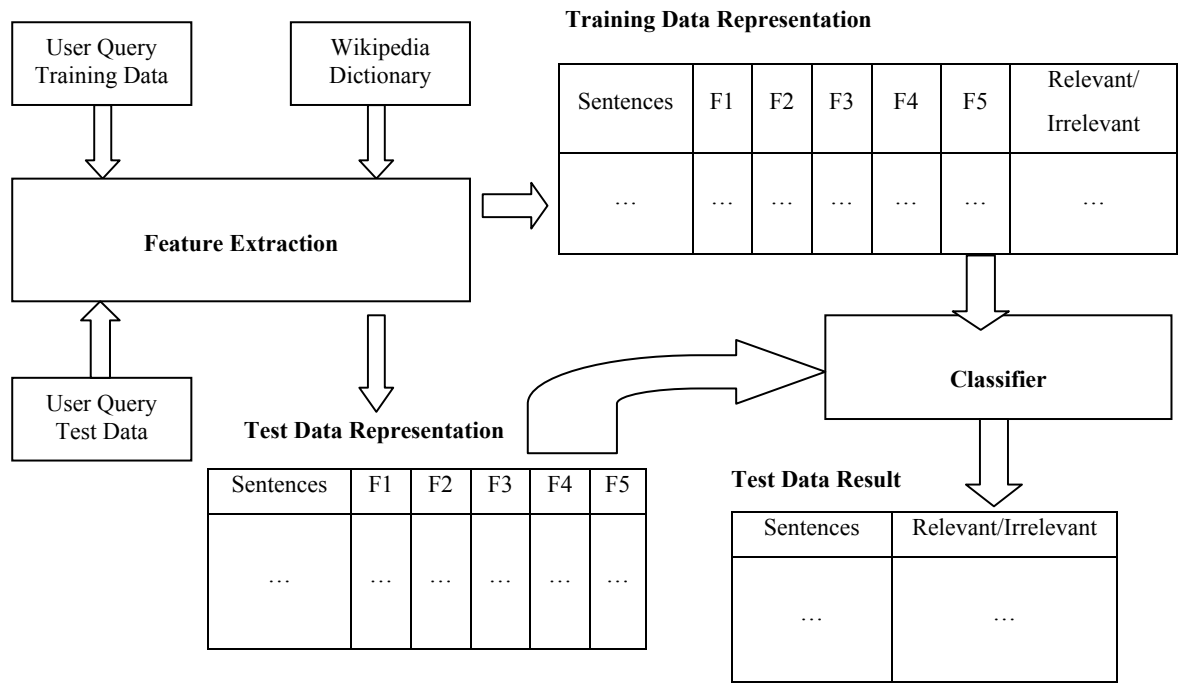

Figure 1. System Architecture for Task 1

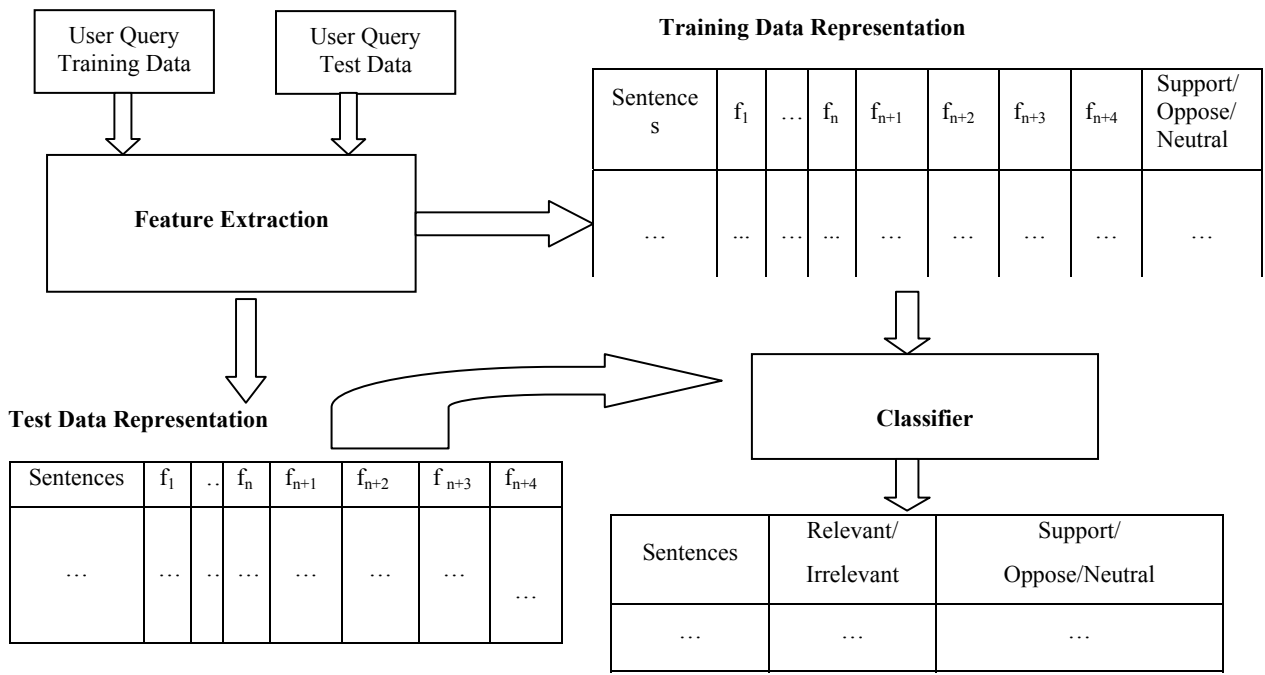

Figure 2. System Architecture for Task 2

Table 1. Performance of the participating systems for Task 1

| Query | Amrita_fire_CEN | JNTUH | Fermi | JU_KS_Group | Techie challangers | SSN_NLP | Amrita_cen | Hua Yang | Jainisha Sankhavara |
|---|---|---|---|---|---|---|---|---|---|
| | | | | | TASK 1 | | | | |
| skincare | 54.54545455 | 62.5 | 78.4 | 48.86363636 | 68.18181818 | 79.54545455 | 48.86363636 | 53.40909091 | 52.27272727 |
| MMr | 87.93103448 | 56.8965517 | 79.31 | 89.65517241 | 87.93103448 | 81.03448276 | 88.88888889 | 84.48275862 | 87.93103448 |
| HRT | 70.83333333 | 38.8888889 | 88.88 | 93.05555556 | 75 | 87.5 | 75.86206897 | 90.27777778 | 91.66666667 |
| Ecig | 71.875 | 57.8125 | 65.62 | 71.875 | 71.875 | 64.0625 | 76.5625 | 46.875 | 54.6875 |
| Vitc | 55.40540541 | 58.1081081 | 72.97 | 63.51351351 | 62.16216216 | 78.37837838 | 60.81081081 | 71.62162162 | 64.86486486 |
| | 68.11804555 | 54.8412097 | **77.036** | **73.39257557** | **73.03000297** | **78.10416314** | **70.19758101** | 69.33324979 | **70.28455866** |

Table 2. Performance of the participating systems for Task 2

| Query | Amrita_fire_CEN | JNTUH | Fermi | JU_KS_Group | Techie challangers | SSN_NLP | Amrita_cen | Hua Yang | Jainisha Sankhavara |
|---|---|---|---|---|---|---|---|---|---|
| | | | | | TASK 2 | | | | |
| skincare | 56.81818182 | 64.7727273 | 73.8 | 44.31818182 | 62.5 | 0 | 23.86363636 | 46.59090909 | 37.5 |
| MMr | 32.75862069 | 65.5172414 | 44.82 | 32.75862069 | 68.96551724 | 0 | 34.72222222 | 63.79310345 | 46.55172414 |
| HRT | 26.38888889 | 48.6111111 | 54.16 | 22.22222222 | 37.5 | 0 | 43.10344828 | 48.61111111 | 27.77777778 |
| Ecig | 37.5 | 67.1875 | 51.56 | 29.6875 | 60.9375 | 0 | 39.0625 | 60.9375 | 46.875 |
| Vitc | 39.18918919 | 31.0810811 | 50 | 39.18918919 | 32.43243243 | 0 | 32.43243243 | 50 | 31.08108108 |
| | 38.53097612 | **55.4339322** | **54.868** | 33.63514278 | **52.46708993** | 0 | 34.63684786 | **53.98652473** | 37.9571166 |

After release of test data by the organizers, we run our system on the test data and send the result files along with the complete system to the organizers. They evaluated the results using the traditional percentage accuracy and published the results which were sent to us through e-mail.

We have shown the officially published results of task 1 and task 2 for the 9 participating teams in Table 1 and Table 2 respectively. The results shown in red bold font are the performances of top systems participated in the tasks.

Out of the 9 participants, our system (JU_KS_Group) achieves the third highest average accuracy for task 1, i.e. 73.39257557%. We can evaluate the results for task 1 in a different angle. It is evident from Table 1 that our system performs better for 3 queries out of 5 queries whereas the system SSN_NLP with the best average accuracy (78.10%) performs better for 2 queries out of five queries. The main reason for my system giving better results for task 1 is the use of two novel features, noun matching and neighborhood matching.

For the task 2, our system achieves an average accuracy of 33.63514278%. For the task 2, our system achieves relatively poor performance. One of the reasons of getting poor performance for task 2 is that we have considered "neutral" class along with other two classes "oppose" and "support" while classifying the relevant sentences. It is evident from the training data that only the irrelevant sentences in the training data were assigned the "neutral" class. Actually the task2 was to classify the relevant sentences into two categories-"Support" and "oppose", but we have mistakenly considered the task2 as 3-class problem instead of 2-class problem. We are working to improve our proposed methods so that our systems can perform more accurately for both the tasks.

## 4. CONCLUSION

There has been a dearth of proper searching systems for medical queries and our work on the CHIS tasks put us on the path to filling this void. The methodology we used can be improved on and innovated with to create a novel searching method for not only medical queries, but any specific search queries of any field. What we have done, and our continuing to improve on, is a logical way of searching through data which is already available to the public. We sincerely believe that through machine learning and natural language processing, the future of online searching can be achieved; and have tried to contribute towards this goal through our paper. And that this will especially be of use in the medical field.

For future work, we would incorporate a word sense disambiguation module to disambiguate the query words. We hope that our system will give more accurate results for task 2 if we consider classification of relevant sentences as 2-class problem ("support" and "oppose") instead of considering it as the 3-class ("support", "oppose" and "neutral") problem that we did during the contest.

ACKNOWLEDGMENTS

We would like to thank the Forum for Information Retrieval Evaluation (FIRE) 2016, ISI Kolkata, for providing us the tasks and datasets for C.H.I.S.